\documentclass[preprint,5p,times,twocolumn]{elsarticle}

\usepackage{amssymb}
\usepackage{amsmath}

\usepackage{physics}

\journal{Optics \& Laser Technology}

\begin{document}

\begin{frontmatter}



\title{Anomalous Second Harmonic Generation of Twisted Gaussian Schell Model Beams}




\author[uff]{M. Gil de Oliveira}
\ead{marcosgil@id.uff.br}
\author[uff]{A. L. S. Santos Junior }
\author[uff]{A. C. Barbosa}
\author[uff]{B. Pinheiro da Silva}
\author[ufsc]{G. H. dos Santos}
\author[miro,collao]{G.~Ca\~{n}as}
\author[ufsc]{P. H. Souto Ribeiro}
\ead{p.h.s.ribeiro@ufsc.br}
\author[miro,uc]{S. P. Walborn}
\ead{swalborn@udec.cl}
\author[uff]{A. Z. Khoury}
\ead{azkhoury@id.uff.br}

\affiliation[uff]{organization={Instituto de Fisica, Universidade Federal Fluminense},
             city={Niteroi},
             state={Rio de Janeiro},
             country={Brazil}}

\affiliation[ufsc]{organization={Instituto de Fisica, Universidade Federal de Santa Catarina},
             city={Florianopolis},
             state={Santa Catarina},
             country={Brazil}}

\affiliation[miro]{organization={Millennium Institute for Research in Optics, Universidad de Concepcion},
             city={Concepcion},
             state={Bio-Bio},
             country={Chile}}

\affiliation[collao]{organization={Departamento de Fisica, Universidad del Bio-Bio, Collao},
             city={Concepcion},
             state={Bio-Bio},
             country={Chile}}

\affiliation[uc]{organization={Departamento de Fisica, Universidad de Concepcion},
             city={Concepcion},
             state={Bio-Bio},
             country={Chile}}

\begin{abstract}
We investigate theoretically and experimentally the optical second harmonic generation (SHG) with a twisted Gaussian Schell model (TGSM) beam as the fundamental field. We use Type-II phase matching and analyze the cross spectral density (CSD) of the SHG output beam when the input fundamental is prepared with a TGSM structure. 
We analyze two synthetization methods for preparing the TGSM fundamental beam and we find that for one
method the SHG is also a TGSM beam. For the other method, we find that the SHG is not a TGSM beam and
presents an anomalous CSD possessing a dip instead of a peak in the transverse spatial structure. 
Moreover, we show that the dip depth is directly related to the twisted phase parameter, being absent 
for a non twisted GSM beam. Our results show that the SHG from a fundamental TGSM beam can result in a
doubled frequency TGSM or in a non-TGSM beam depending on the synthetization method.
\end{abstract}


\end{frontmatter}

\section{Introduction}

Optical fields are partially coherent for practically all natural sources like the sun, stars and other thermal sources. One of the important features of laser light is the high degree of coherence, which is pursued by the constructors by means of high quality cavities and frequency stabilization techniques. However, there has been an increasing interest in some special partially coherent light fields. The Twisted Gaussian Schell-model (TGSM) beams are attracting a high deal of interest due to several relevant applications \cite{Cai2022,ismail2017,ismail2020,Hutter21,Zhang:19,Ma:17}. TGSM beams were invented in 1992 by Simon and Mukunda by introducing a position dependent ``twist" phase in the correlation function of GSM beams \cite{simon93} and experimentally realized by Friberg et. al. \cite{friberg94}. The twist phase is different from the orbital angular momentum of light because they can only exist in partially coherent beams, while orbital angular momentum is a property of a highly coherent light mode. 

Besides TGSM beams, other classes of TGSM beams where created and implemented in the laboratory. Some examples are the twisted Laguerre Gaussian Schell-model (TLGSM) \cite{Peng:18}, twisted Hermite GSM (THGSM) beams \cite{Xiaofeng2022}, ring-shaped twisted Gaussian Schell-model array (RTGSMA) \cite{Zheng2020}, and twisted vortex Gaussian Schell-model (TVGSM) beams \cite{Stahl:18}. There are interesting applications for all these beams and we are particularly concerned with TGSM beams. We are motivated by applications like optical communication through atmospheric and underwater turbulence \cite{Cai2006,Wang:10,Wang:12,Peng:17,Zhou:18}, in resisting coherence induced depolarization, in overcoming the classical Rayleigh limit \cite{Tong:12}, to control the coherence of optical solitons \cite{Ponomarenko2001}, to boost entanglement in photon pairs \cite{Hutter20,Hutter21}, and in stimulated parametric down-conversion \cite{dos2022phase}.

TGSM beams are not available in nature and they need to be prepared in the laboratory. Moreover, it is always convenient generating TGSM beams starting from a laser source, due to the high photon number occupation per mode provided by lasers. Therefore, generation and manipulation of these special beams is an important research topic nowadays \cite{friberg94,wang19,Stahl:18,tian20,zhang2021generating,Liu:22, wang2022experimental,canas2022}. Among the available approaches for generating TGSM  beams, there are approaches based on conversion of GSM beams like the one  by Friberg et al. \cite{friberg94}, who employed an experimental setup consisting of a combination of six-cylindrical lenses and a variable-coherence anisotropic GSM source and the one by Wang et al., which converted an anisotropic GSM beam into a TGSM beam using a set of three cylindrical lenses \cite{wang19}. Other methods are based on mode decompositions. For instance, TGSM beams were created by implementing the continuous coherent beam integral function in a discrete form \cite{tian20,canas2022} and in Ref. \cite{tian20}, a Laguerre-Gaussian mode decomposition was used, based on the theory introduced in \cite{simon93b,sundar93}. In a more recent realization, TGSM beams with controllable twist phase were produced with an incoherent superposition of random modes obeying Gaussian statistics \cite{wang2022experimental}. More sophisticated classes of partially coherent beams can also be generated with these approaches. For instance the radially polarized twisted partially coherent vortex  (RPTPCV) was prepared by Liu et al. \cite{Liu:22}. \\

In nonlinear optics, there has been some studies concerning the propagation characteristics of TGSM beams in non-linear Kerr media \cite{hu_influence_2021}. The theory for treating TGSM in nonlinear interactions is being developed and has been recently advanced by Zheltikov et al. \cite{zheltikov_modulation_2023}. Experiments with TGSM beams in nonlinear interactions were also reported, as for instance Ref. \cite{dos2022phase}, where phase conjugation is demonstrated in stimulated parametric down-conversion.

In this paper, we study the nonlinear interaction of a TGSM beam in a second harmonic generation (SHG) process. Due to the nonlinear coupling between the optical fields, the equation that governs this process couples correlation functions of different orders. We show that different methods for generating TGSM lead to different fourth order correlation functions. Therefore, the output up-converted beam depends upon which generation method is used, as well as the SHG experimental setup.

This article is structured as follows: in section \ref{sec:TGSM Beams} we briefly review the properties of the TGSM beam and discuss two different synthesizing methods. In section \ref{sec:Second Harmonic Generation with TGSM beams} we derive theoretical predictions for the output of a SHG process when the fundamental beams are TGSM, considering different experimental setups and TGSM generation techniques. In section \ref{sec:experimental_results}, we present experimental results that corroborate our theoretical findings. Finally, in section \ref{sec:conclusion}, we draw our conclusions.

\section{TGSM Beams}
\label{sec:TGSM Beams}

For a partially coherent source, the electric field is a random variable, for which we cannot assign a field value, being only able to calculate averages. Let us write the Fourier component at frequency $\omega$ of the electric field as $\mathbf{E}(\mathbf{r},\omega) = \mathcal{E}(\mathbf{r}) e^{i \mathbf{k} \cdot \mathbf{r}} \hat{\mathbf{e}}$, where $\mathbf{k}$ is the wavevector and $\hat{\mathbf{e}}$ is the polarization vector, which is assumed to be fixed. Then, the cross-spectral density (CSD) function, which can be used to characterize the partially coherent source, is defined as
\begin{equation}
    \Gamma(\mathbf{r},\mathbf{r}^\prime) = \expval{ \mathcal{E}(\mathbf{r}) \mathcal{E}^*(\mathbf{r}^\prime) },
\end{equation}
where $\mathbf{r} = (x,y)$ is the transverse position and $\expval{ \cdot }$ denotes the ensemble average.

The TGSM \cite{simon93,simon93b,sundar93,simon98} beam describes a class of partially coherent beams that have a Gaussian intensity profile, a Gaussian degree of coherence function, and that include a novel twist-phase $\mu$ that can give them non-zero optical angular momentum. In the focal plane, their CSD function is given by
\begin{equation}
\label{eq:general_csd_tgsm}
    \Gamma(\mathbf{r},\mathbf{r}^\prime) = \abs{A}^2 T(\mathbf{r},\mathbf{r}^\prime),
\end{equation}
where $A$ is a field amplitude and
\begin{equation}
\label{eq:tgsm}
\begin{aligned}
    T(\mathbf{r},\mathbf{r}^\prime) &= T(\mathbf{r},\mathbf{r}^\prime;w,\delta,k,\mu) \\ &= e^{-\frac{r^2 + r^{\prime 2}}{4 w^2}-\frac{(\mathbf{r} - \mathbf{r}^{\prime})^2}{2 \delta^2}-ik\mu \mathbf{r} \wedge \mathbf{r}^\prime}.
\end{aligned}
\end{equation}
Here, $\mathbf{r} \wedge \mathbf{r}^\prime=x y^\prime - x^\prime y$, $2 w$ is the beam waist and $\delta$ is the transverse coherence length. The twist phase $\mu$ satisfies the condition $k |\mu| \leq 1/\delta^2$ \cite{simon93,simon98}. Therefore, it is useful to introduce the dimensionless twist phase
\begin{equation}
    \tau = k \mu \delta^2,
\end{equation}
which satisfies $\abs{\tau} \le 1$. We will also introduce the dimensionless transverse coherence length, defined  by
\begin{equation}
    q = \frac{\delta}{\sqrt{2} w}.
\end{equation}

\subsection{TGSM generation methods}
As in previous work, we will use phase-randomized modes in order to produce the TGSM beams \cite{tian20,canas2022}. In practice, these modes can be produced using a sequence of phase masks imprinted on a spatial light modulator or similar device. This technique is based on the properties of the stochastic field
\begin{equation}
    \label{eq:stochastic_field}
    \Psi(\mathbf{r}) = \sum_n \sqrt{\lambda_n}  K_n(\mathbf{r}) e^{i\phi_n},
\end{equation}
where $\{ K_n\}$ is a family of coherent modes, $\lambda_n$ are a set of corresponding weights and $\phi_n$ are random independent phases that are uniformly distributed in $[0,2\pi)$. By using
\begin{equation}
    \left< e^{i\left(\phi_{m} - \phi_{n}\right)}  \right> = \delta_{mn},
\end{equation}
one can see that this field has a correlation function given by
\begin{equation}
    \label{eq:avg_stochastic_field}
    \left< \Psi(\mathbf{r})  \Psi^*(\mathbf{r}^\prime) \right> = \sum_n \lambda_n K_n(\mathbf{r}) K_n^*(\mathbf{r}^\prime).
\end{equation}
Therefore, if we find coherent modes $K_n$ and weights $\lambda_n$ such that
\begin{equation}
\label{eq:universal_decomposition}
    T(\mathbf{r},\mathbf{r}^\prime) \approx \sum_n \lambda_n K_n(\mathbf{r}) K_n^*(\mathbf{r}^\prime),
\end{equation}
we conclude that this stochastic field \eqref{eq:stochastic_field} is a TGSM.

In practice, the averaging in \eqref{eq:avg_stochastic_field} is done by producing many realizations of the fields \eqref{eq:stochastic_field} with different randomly chosen phases, and summing the output results, either at the detection stage or posteriori.
We will analyze two possible choices of modes. The first one is based on the decomposition
\begin{equation}
    \label{eq:continuous_decompostion}
    \begin{aligned}
        T\left(\mathbf{r},\mathbf{r}^\prime\right) &= \int d^2 \mathbf{v} \ p(\mathbf{v}) DG(\mathbf{r},\mathbf{v}) DG^*(\mathbf{r}^\prime ,\mathbf{v}),
    \end{aligned}
\end{equation}
where
\begin{equation}
    \begin{aligned}
        DG(\mathbf{r},\mathbf{v}) = & \exp\left[ - \frac{w^2}{2\alpha w^2+1}\left( \frac{\mathbf{r}}{2w^2} + \alpha \mathbf{r} - \alpha \mathbf{v} \right)^2 \right] \\ & \times \exp\left[-ik\mu(xv_y-yv_x)\right],
    \end{aligned}
\end{equation}
are dislocated Gaussians (DG), and
\begin{equation}
    p(\mathbf{v}) = \frac{\alpha}{\pi}\exp \left( - \frac{\alpha v^2}{2\alpha w^2+1} \right)
\end{equation}
is a weight function with
\begin{equation}
    \alpha = \frac{1 + \sqrt{1-\tau^2}}{\delta^2}.
\end{equation}

By truncating and discretizing, we may put expression \eqref{eq:continuous_decompostion} in the form of equation \eqref{eq:universal_decomposition}, by choosing $K_n(\mathbf{r}) = DG(\mathbf{r},\mathbf{v}_n)$, where $\mathbf{v_n}$ is a point in the discretized grid, and 
\begin{equation}
\label{eq:lambda_n-p(v)}
    \lambda_n = \Delta A p(\mathbf{v}_n)\,, 
\end{equation}
where $\Delta A$ is the discretization area.

Another possible decomposition is obtained by using the Laguerre-Gauss basis, with modes given by
\begin{equation}
	LG_{pl}(\mathbf{r}) = \mathcal{N}_{pl} \!
	\left(2\tilde{r}^2\right)^{\!\!\frac{|l|}{2}} \!
	L_{p}^{|l|}\!\left(2\tilde{r}^2\right)
	e^{-\tilde{r}^2}\!e^{il\phi},
\label{eq:modefunctions}
\end{equation}
where $w_0$ is the beam's waist, $\tilde{r} = r/w_0$ and 
\begin{equation}
    \mathcal{N}_{pl} = \frac{1}{w_0}\sqrt{\frac{2}{\pi}}\sqrt{\frac{p!}{(p+|l|)!}}
\end{equation}
is the normalization constant. 
As shown in \cite{simon93b,sundar93}, we then have 
\begin{equation}
\label{eq:laguerre_decomposition}
    \sum_{p,l} \lambda_{pl}  LG_{pl}(\mathbf{r}) LG_{pl}^*(\mathbf{r}^\prime) = T\left(\mathbf{r},\mathbf{r}^\prime;w,\delta,\mu,k\right),
\end{equation}
where the weights are given by
\begin{equation}
\label{eq:weight_laguerre}
\begin{aligned}
    \lambda_{pl}&=\frac{\pi}{2} w_0^2(1-\xi) \left(\frac{1+\tau}{1-\tau} \right)^{l/2} \xi^{\frac{|l|}{2}+p} \\ &=\lambda_{00} \left(\frac{1+\tau}{1-\tau} \right)^{l/2} \xi^{\frac{|l|}{2}+p} ,
\end{aligned}
\end{equation}
with parameters
\begin{equation}
    \frac{w_0}{2w} = \frac{q}{\left( q^4+2q^2+\tau^2\right)^{1/4}},
\end{equation}
and
\begin{equation}
    \xi=\frac{1+q^2 - \sqrt{q^4+2q^2+\tau^2}}{1+q^2 + \sqrt{q^4+2q^2+\tau^2}}.
\end{equation}

When $\abs{\tau} \to 1$, equation \eqref{eq:weight_laguerre} becomes undefined, but one may take the limit, obtaining
\begin{equation}
    \lim_{\abs{\tau}\to 1} \lambda_{pl} = \begin{cases}
        \dfrac{\lambda_{00} \delta_{p0}}{(1+q^2)^{\abs{l}}} & \tau l \ge 0 \\
        0 & \tau l < 0
    \end{cases}
\end{equation}

\section{Second Harmonic Generation with TGSM beams}
\label{sec:Second Harmonic Generation with TGSM beams}
\begin{figure}
    \includegraphics[width=\linewidth]{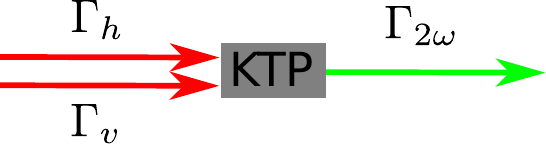}
    \caption{Type-II SHG in a Potassium Titanyl Phosphate (KTP) crystal. The inputs are partially coherent beams, resulting in a partially coherent output beam.}
    \label{fig:shgtgsm}
\end{figure}
Let us consider now SHG in a non-linear crystal, as illustrated in Fig. \ref{fig:shgtgsm}. For a thin crystal, the second harmonic field $\mathcal{E}_{2\omega}$ is given by \cite{pinheiro22,da2022observation}
\begin{equation}
    \mathcal{E}_{2\omega}(\mathbf{r}) = ig \mathcal{E}_{h}(\mathbf{r}) \mathcal{E}_{v}(\mathbf{r}),
\end{equation}
where $g$ is a real and positive coupling constant and $\mathcal{E}_{h(v)}(\mathbf{r})$ is the horizontal (vertical) component of the fundamental beam(s). The CSD function is then
\begin{equation}
\label{eq:complete_csd_sh}
    \begin{aligned}
        \Gamma_{2 \omega}\left(\mathbf{r}, \mathbf{r}^\prime \right) &= \left< \mathcal{E}_{2\omega}(\mathbf{r}) \mathcal{E}^*_{2\omega}(\mathbf{r}^\prime) \right> \\ &= g^2 \left< \mathcal{E}_{h}(\mathbf{r}) \mathcal{E}_{v}(\mathbf{r}) \mathcal{E}_{h}^*(\mathbf{r}^\prime) \mathcal{E}_{v}^*(\mathbf{r}^\prime) \right>.
    \end{aligned}
\end{equation}

Thus, the CSD function for the second harmonic is expressed in terms of a \textit{fourth order} correlation, involving both the vertical and horizontal fields. If we have a single incident beam with diagonal polarization, then the spatial field profiles of the horizontal and vertical polarization components will be the same: $\mathcal{E}_{h}(\mathbf{r}) = \mathcal{E}_{v}(\mathbf{r}) = \mathcal{E}(\mathbf{r})$. Note that the knowledge of the CSD
\begin{equation}
    \Gamma_{h(v)}\left(\mathbf{r}, \mathbf{r}^\prime \right) = \left< \mathcal{E}_{h(v)}(\mathbf{r}) \mathcal{E}^*_{h(v)}(\mathbf{r}^\prime) \right>,
\end{equation}
 is not sufficient to determine \eqref{eq:complete_csd_sh}, in general.

We thus conclude that proper evaluation of \eqref{eq:complete_csd_sh} requires a more detailed specification of the statistical properties of $\mathcal{E}_{h(v)}(\mathbf{r})$.

\subsection{Independent Polarizations}

From the theoretical perspective, the simplest possible situation is when the $h$- and $v$-polarized beams are prepared independently. In this case, the average in \eqref{eq:complete_csd_sh} involves averaging over two independent random fields, and thus factors into a product of the CSD function for each polarization, giving simply
\begin{equation}
\label{eq:csd_sh_independent}
    \begin{aligned}
        \Gamma_{2 \omega}\left(\mathbf{r}, \mathbf{r}^\prime \right) &= g^2 \Gamma_{h}\left(\mathbf{r}, \mathbf{r}^\prime \right) \Gamma_{v}\left(\mathbf{r}, \mathbf{r}^\prime \right).
    \end{aligned}
\end{equation}

If the CSDs $\Gamma_{h(v)}$  are those of TGSM beams, described by equation \eqref{eq:general_csd_tgsm}, then $\Gamma_{2\omega}$ is given by
\begin{equation}
    \begin{aligned}
        \frac{\Gamma_{2 \omega}\left(\mathbf{r}, \mathbf{r}^\prime \right)}{g^2 \abs{A_h}^2 \abs{A_v}^2} &= T(\mathbf{r},\mathbf{r}^\prime;w_h,\delta_h,k,\mu_h)T(\mathbf{r},\mathbf{r}^\prime;w_v,\delta_v,k,\mu_v) \\ &= T\left( \mathbf{r}, \mathbf{r}^\prime; w_{2\omega}, \delta_{2\omega}, 2k, \mu_{2\omega}\right).
    \end{aligned}
\end{equation}
Thus, the output SHG beam is also a TGSM beam, whose beam parameters are related to those of the input fundamental beams by
\begin{equation}
    \frac{1}{X_{2\omega}}= \frac{1}{X_{h}}+ \frac{1}{X_{v}},      
    \label{eq:X}
\end{equation}
for $X=w,\delta$ and
\begin{equation}
    \mu_{2\omega} = \frac{\mu_h + \mu_v}{2}.
\end{equation}
From this, we see that the constraint on the twist phase is satisfied, as these relations lead directly to
\begin{equation}
    \mu_{2\omega} \leq \frac{1}{2k \delta_{2\omega}^2}.
\end{equation}

\subsection{Diagonal polarization}

If instead of independently prepared input beams, we have a single incident beam with diagonal polarization, then the spatial field profiles of the horizontal and vertical polarization components will be equal: $\mathcal{E}_{h}(\mathbf{r}) = \mathcal{E}_{v}(\mathbf{r}) = \mathcal{E}_{\omega}(\mathbf{r})$, and therefore,
\begin{equation}
    \label{eq:equal_csd_sh}
    \Gamma_{2 \omega}\left(\mathbf{r}, \mathbf{r}^\prime \right) = g^2\left< \left[\mathcal{E}_{\omega}(\mathbf{r}) \mathcal{E}^*_{\omega}(\mathbf{r}^\prime)\right]^2\right>.
\end{equation}

Let us assume that our input field has the form $\mathcal{E}_\omega(\mathbf{r}) = A \Psi(\mathbf{r})$, where $A$ is a field amplitude and $\Psi(\mathbf{r})$ is given by \eqref{eq:stochastic_field}. Then, the second harmonic CSD is
\begin{equation}
    \label{eq:csd_sh_partial}
    \Gamma_{2 \omega}\left(\mathbf{r}, \mathbf{r}^\prime \right) = g^2 \abs{A}^4 \left< \left[\Psi(\mathbf{r})  \Psi^*(\mathbf{r}^\prime)\right]^2 \right>.
\end{equation}
In order to evaluate this expression, we need to use the property
\begin{equation}
    \label{eq:average_four_phases}
    \begin{aligned}
        \left< e^{i\left(\phi_{m} + \phi_{n} - \phi_{j} - \phi_{k}\right)}  \right> &= \begin{cases}
            1, & m = j \text{ and } n = k \\
            1, & m = k \text{ and } n = j \\
            0, & \text{otherwise}
        \end{cases} \\
        &= \delta_{mj}\delta_{nk}\left( 1 - \delta_{mn}\right) + \delta_{mk}\delta_{nj}.
    \end{aligned} 
\end{equation}
Using this result, we obtain
\begin{equation}
    \label{eq:shg_correlation_rp}
    \begin{aligned}
        \frac{\Gamma_{2\omega}\left(\mathbf{r},\mathbf{r}^\prime \right)}{g^2 \abs{A}^4} &= \left< \left[\Psi(\mathbf{r})  \Psi^*(\mathbf{r}^\prime)\right]^2 \right> \\ &= 2 T^2\left(\mathbf{r},\mathbf{r}^\prime\right) - \gamma_1(\mathbf{r}, \mathbf{r}^\prime)
    \end{aligned}
\end{equation}
where
\begin{equation}
     \gamma_1(\mathbf{r}, \mathbf{r}^\prime) = \sum_n \left[\lambda_n K_n(\mathbf{r}) K^*_n(\mathbf{r}^\prime)\right]^2.
\end{equation}

It is interesting to notice that, when using the dislocated Gaussian decomposition, the last term $\gamma_1$ is negligible. One can see this by recalling Eq. \eqref{eq:lambda_n-p(v)}, and taking the continuous limit:
\begin{equation}
    \begin{aligned}
        \gamma_1(\mathbf{r}, \mathbf{r}^\prime)&=\Delta A^2 \sum_n \left[p(\mathbf{v}_n) DG(\mathbf{r},\mathbf{v}_n) DG^*(\mathbf{r}^\prime ,\mathbf{v}_n)\right]^2 \\
         &\xrightarrow[\Delta A \to 0]{} \Delta A\int d^2 \mathbf{v} \left[ p(\mathbf{v}) DG(\mathbf{r},\mathbf{v}) DG^*(\mathbf{r}^\prime ,\mathbf{v}) \right]^2 \\
         &\xrightarrow[\Delta A \to 0]{} 0.
    \end{aligned}
\end{equation}
Then, the CSD function for the second harmonic reduces to
\begin{equation}
    \label{eq:shg_correlation_rp_continuous.}
    \begin{aligned}
        \frac{\Gamma_{2\omega}\left(\mathbf{r},\mathbf{r}^\prime \right)}{g^2 \abs{A}^4} &= T^2\left(\mathbf{r},\mathbf{r}^\prime; w,\delta,k,\mu\right) \\ &= T\left(\mathbf{r},\mathbf{r}^\prime; w/\sqrt{2},\delta/\sqrt{2},2k,\mu\right),
    \end{aligned}
\end{equation}
corresponding to a TGSM beam.
\par
\begin{figure}
    \centering
    \includegraphics[width=\linewidth]{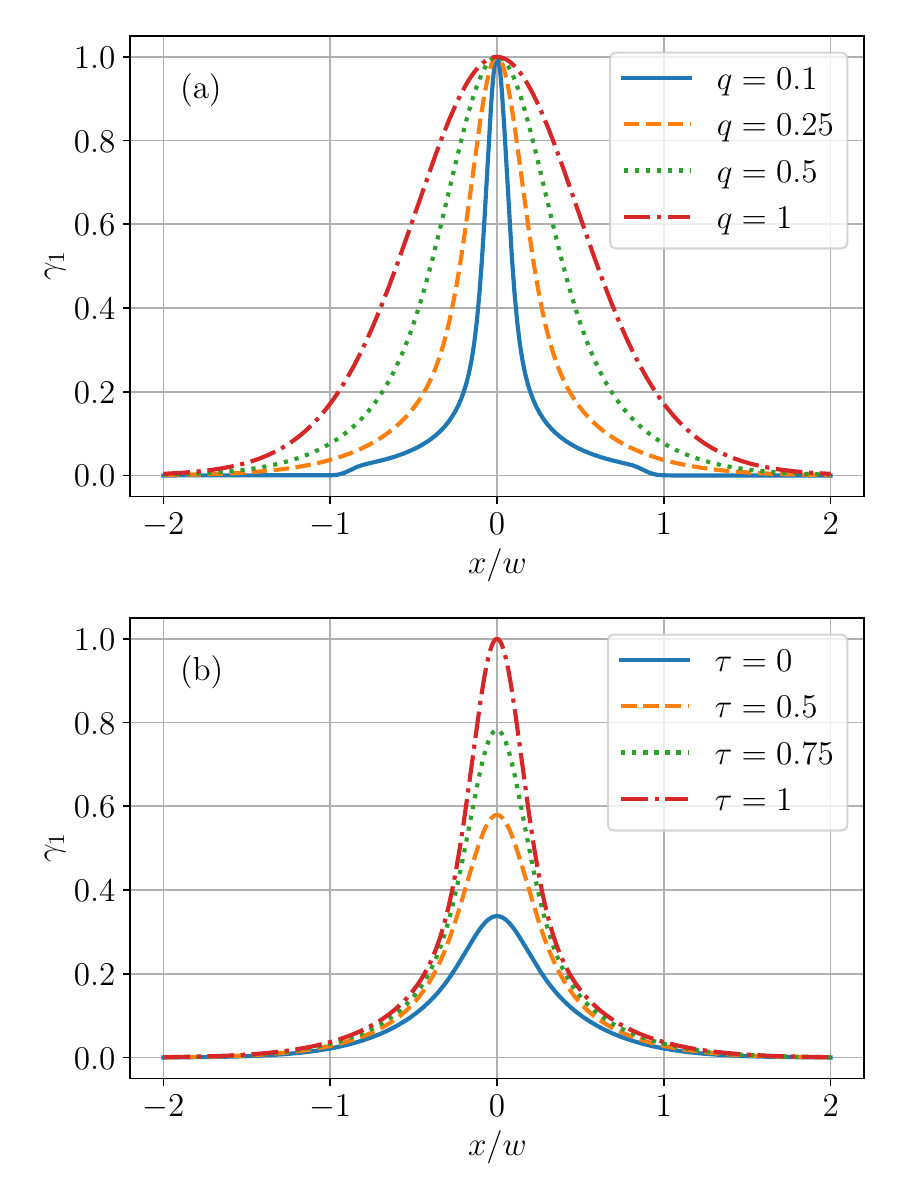}
    \caption{$\gamma_1$ evaluated at $\mathbf{r} = \mathbf{r}^\prime = (x,0)$. On the top, we fix $\tau = 1$, and vary $q$. At the bottom, we fix $q = 1/4$ and vary $\tau$.}
    \label{fig:peaks}
\end{figure}

On the other hand, $\gamma_1$ has an important effect when using the Laguerre-Gauss decomposition \eqref{eq:laguerre_decomposition}. As may be seen from its numerical summation, plotted in Figure \ref{fig:peaks}, the correction to the intensity $\gamma_1(\mathbf{r},\mathbf{r})$ presents a single maximum at $\mathbf{r} = \mathbf{0}$. The value of this maximum can be easily calculated because, at the origin, only the modes with $l = 0$ contribute, giving us
\begin{equation}
    \label{eq:gamma_shg_peak}
    \begin{aligned}
        \gamma_1(\mathbf{0},\mathbf{0}) &= \left( \frac{\pi}{2} w_0^2 \right)^2 (1-\xi)^2 \sum_{p=0}^\infty  \abs{LG_{p0}(\mathbf{0})}^4\xi^{2p} \\ &= \sqrt{1 - \frac{1 - \tau^2}{\left(1 + q^2 \right)^2}}.
    \end{aligned}
\end{equation}

We see that, for maximum twist phase $\abs{\tau} = 1$, the peak attains a maximum value of $1$, which in independent of the normalized coherence length $q$. For all other values of $\tau$, the peak increases as $q$ decreases. Figure \ref{fig:peaks} also reveals that, when $q$ increases, the width of the peak also increases, and its shape begins to resemble a Gaussian, although it strongly deviates from it when $q \ll 1$.

\begin{figure}
    \centering
    \includegraphics[width=\linewidth]{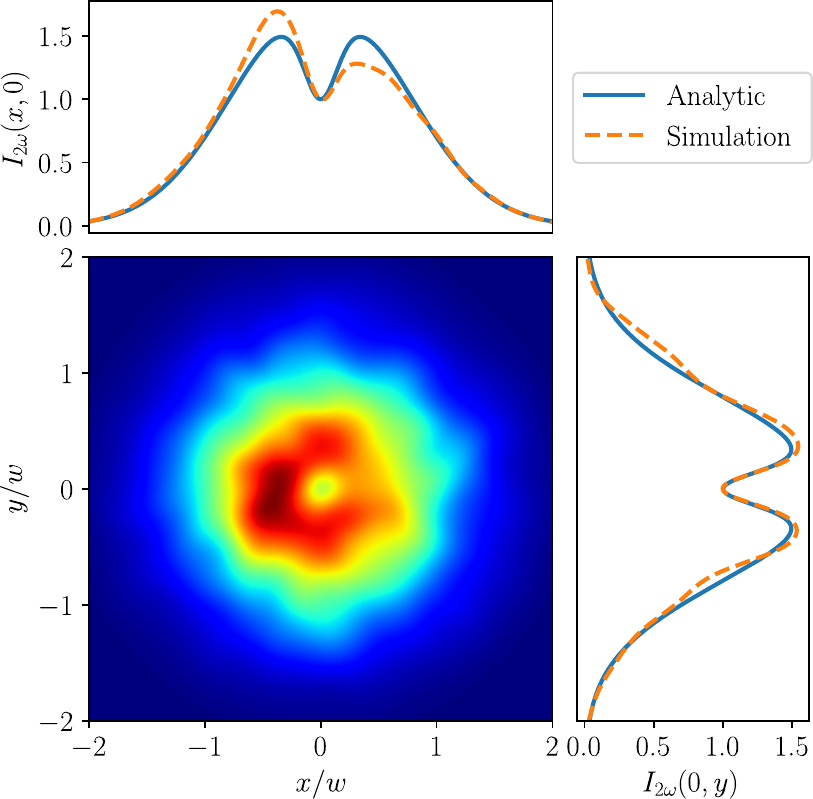}
    \caption{Center: simulated intensity profile $I_{2\omega}(\mathbf{r}) = \Gamma_{2\omega}\left(\mathbf{r},\mathbf{r} \right) / g^2 \abs{A}^4$. The simulation is produced by averaging over $300$ different realizations of the stochastic field \eqref{eq:stochastic_field}. We used $q = 1/4$ and $\tau = 1$. On the top and right we see cuts of the simulated intensity profile through the lines $y=0$ and $x = 0$, respectively. We also display the analytical result, which corresponds to expression \eqref{eq:shg_correlation_rp} evaluated over these lines.}
    \label{fig:simulation_shg}
\end{figure}

When $\gamma_1(\mathbf{r},\mathbf{r})$ is thin and peaked, which corresponds to $q \ll 1$ and $\abs{\tau} \approx 1$, we expect that it will pierce the Gaussian in \eqref{eq:shg_correlation_rp}, provoking a dip at the center of the intensity profile. This effect can be clearly seen in the simulation presented in Figure \ref{fig:simulation_shg}.

Thus, we note that in case of the Laguerre-Gauss decomposition a different correlation function is expected in the second harmonic when compared to the dislocated gaussian decomposition, corresponding to a beam that is not a TGSM beam. In section \ref{sec:experimental_results}, we will present experimental verifation of these theoretical predictions.

\subsection{Correlated polarizations}

A third interesting case is  when each polarization in the fundamental beams are prepared with conjugate modes. More specifically, if we set $\mathcal{E}_h(\mathbf{r}) = A_h \Psi(\mathbf{r})$ and $\mathcal{E}_v(\mathbf{r}) = A_v \Psi^*(\mathbf{r})$, where, once again, $\Psi$ is the stochastic field \eqref{eq:stochastic_field}, then
\begin{equation}
\label{eq:plus_mu}
    \Gamma_{h}(\mathbf{r},\mathbf{r}^\prime) = \abs{A_h}^2T(\mathbf{r},\mathbf{r}^\prime;w,\delta,k,\mu),
\end{equation}
\begin{equation}
    \Gamma_{v}(\mathbf{r},\mathbf{r}^\prime) = \abs{A_v}^2T(\mathbf{r},\mathbf{r}^\prime;w,\delta,k,-\mu)
\end{equation}
and
\begin{equation}
    \Gamma_{2\omega}\left( \mathbf{r}, \mathbf{r} ^\prime \right) = g^2 \abs{A_h}^2\abs{A_v}^2 \expval{\abs{\Psi(\mathbf{r})}^2 \abs{\Psi(\mathbf{r}^\prime)}^2}.
\end{equation}
By once more applying property \eqref{eq:average_four_phases}, we get
\begin{equation}
    \label{eq:shg_correlation_rp2}
    \begin{aligned}
        \frac{\Gamma_{2\omega}\left( \mathbf{r}, \mathbf{r} ^\prime \right)}{g^2 \abs{A_h}^2\abs{A_v}^2}  = \ & T^2\left( \mathbf{r}, \mathbf{r} ^\prime \right) + T\left( \mathbf{r}, \mathbf{r}\right)T\left( \mathbf{r}^\prime , \mathbf{r} ^\prime \right) - \gamma_2(\mathbf{r},\mathbf{r}^\prime),
    \end{aligned}
\end{equation}
with $T$ given by equation \eqref{eq:plus_mu} and
\begin{equation}
    \gamma_2(\mathbf{r},\mathbf{r}^\prime) = \sum_n \lambda_n^2 \abs{K_n(\mathbf{r})}^2\abs{K_n(\mathbf{r}^\prime)}^2.
\end{equation}

In some aspects, this expression is similar to \eqref{eq:shg_correlation_rp}: $\gamma_2$ also goes to zero in the case of dislocated Gaussians, but has a relevant contribution for the Laguerre-Gaussians. In the latter case, the exact same dip in intensity predicted for diagonally-polarized fundamental beam is also present, since evaluating \eqref{eq:shg_correlation_rp} and \eqref{eq:shg_correlation_rp2} at $\mathbf{r} = \mathbf{r}^\prime$ gives the same result, apart from a multiplicative factor.

Nonetheless, there is a stark difference: even in the case of the dislocated Gaussians, we have
\begin{equation}
    \frac{\Gamma_{2\omega}\left( \mathbf{r}, \mathbf{r} ^\prime \right)}{g^2 \abs{A_h}^2\abs{A_v}^2}  = T^2\left( \mathbf{r}, \mathbf{r} ^\prime \right) + T\left( \mathbf{r}, \mathbf{r}\right)T\left( \mathbf{r}^\prime , \mathbf{r} ^\prime \right),
\end{equation}
which is not the correlation function for a TGSM. One should note that this difference will not show up in the intensity profile, but only in the field correlations. To make this explicit, let's suppose that $T$ is a TGSM with very low coherence ($q \ll 1$). Then, for $r \sim w$, we have
\begin{equation}
    T(\mathbf{r},\mathbf{0}) \approx 0,
\end{equation}
but
\begin{equation}
    \frac{\Gamma_{2\omega}(\mathbf{r},\mathbf{0})}{2g^2 \abs{A_h}^2\abs{A_v}^2} \approx T(\mathbf{r},\mathbf{r}) \not \approx 0.
\end{equation}
In words, $\Gamma_{2\omega}$ will carry a residual coherence that does not depend on the coherence length of $\Gamma_{h(v)}$.

We note that the correlation function which is here discussed is also expected in a Stimulated Parametric Down Conversion process, in which an idler field $\mathcal{E}_i$ is produced by a combination of a pump beam $\mathcal{E}_p\left( \mathbf{r}\right) \propto \Psi\left( \mathbf{r}\right)$ and a seed beam $\mathcal{E}_s\propto \Psi\left( \mathbf{r}\right)$, which are combined through the equation
\begin{equation}
    \mathcal{E}_i\left( \mathbf{r}\right) = ig\mathcal{E}_p\left( \mathbf{r}\right) \mathcal{E}_s^*\left( \mathbf{r}\right),
\end{equation}
which is, again, valid within the thin crystal approximation.

\section{Experimental Results}
\label{sec:experimental_results}

\begin{figure*}
    \centering
    \includegraphics[width=\linewidth]{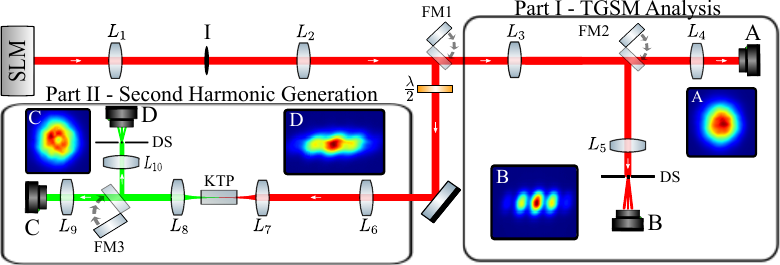}
    \caption{Experimental setup.}
    \label{fig:setup_exp}
\end{figure*}
To test the anomaly in the nonlinear optical conversion of TGSM beams, we experimentally implement Second Harmonic Generation with TGSM beams in the case of diagonally polarized fundamental beam.

The experimental setup can be seen in Figure \ref{fig:setup_exp}. An infrared Gaussian beam with wavelength of $\lambda=1064$ nm and waist of $3.3$mm illuminates the spatial light modulator (SLM) screen. We use this configuration to generate TGSM beams, as described in \cite{tian20,canas2022}. All the beams were produced with a waist of $0.7$mm and a normalized coherence length of $q = 0.4$. We allowed $\tau$ to span the values $0$ and $\pm1$. The intensities were obtained by averaging over 300 realizations of the stochastic field \eqref{eq:stochastic_field}.

As illustrated in the figure, we employ lenses $L_1$ and $L_2$ (with focal length of $ 20$cm) to create a 4f optical system, where the iris (I) is positioned between the lenses to filter the first-order diffraction.

The experiment is divided into two distinct parts. In the first part, we characterize the incident infrared TGSM to ensure that both preparation methods produce the same beam in the fundamental frequency. Due to the field diffraction during propagation, we employ a second 4f optical system, composed of lenses $L_3$ and $L_4$ ($f = 10$cm), to project the SLM field image onto camera A. This direct intensity measurement allows us to determine the parameter $w$. Using the flip mirror 2 (FM2) and an additional 4f system formed by lenses $L_3$ and $L_5$ ($f = 10$cm), we deflect the beam to pass it through a double slit (with spacing of $0.2$mm and aperture of $0.05$mm), resulting in an interference pattern observed by camera B, positioned in the far field of the double slit. The visibility of this pattern is directly related to the coherence length $\delta$, while the twist $\mu$ has the effect of tilting the lobes of this pattern \cite{canas2022}.

For the second part of the experiment, we directed the beam using flip mirror 1 (FM1) to the half-wave plate to prepare the input beam with diagonal polarization. We focused the beam using a 4f optical system consisting of lenses $L_6$ ($f = 20$ cm) and $L_7$ ($f = 5$ cm) on the Potassium Titanyl Phosphate (KTP) crystal to observe its second harmonic at wavelength of $\lambda = 532$nm. It is separated from the fundamental beam using a spectral filter. The 4f system, consisting of lenses $L_8$ ($f = 5$ cm) and $L_9$ ($f = 20$ cm), magnified the beam that could be captured by camera C. We could deflect the second harmonic beam with a flip mirror (FM3) and make it pass through a double slit identical to the previous one. Utilizing another 4f system composed of $L_8$ and $L_{10}$ ($f = 10$ cm), we can observe in the far field the interference pattern of the second harmonic beam through camera D.

\begin{figure*}
    \centering
    \includegraphics[width=\linewidth]{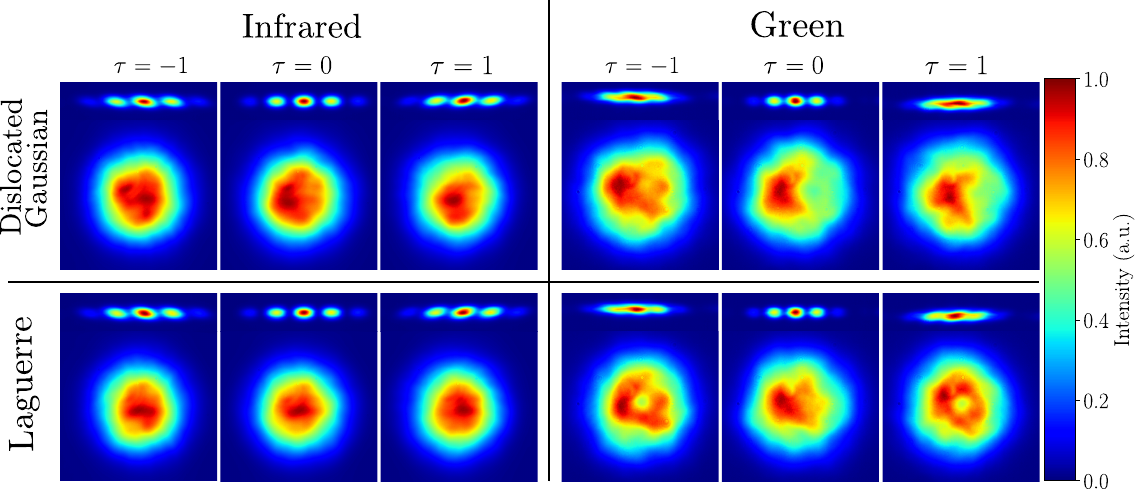}
    \caption{Experimental Results. The top row corresponds to the dislocated Gaussian decomposition, while the bottom one corresponds to the Laguerre decomposition. The left column presents the results of the infrared measurements, while the right one corresponds to measurements performed after the second harmonic generation (green). On top of each direct measurement, there is the corresponding double slit interference pattern. }
    \label{fig:experimental_results}
\end{figure*}

Our results are shown in Figure \ref{fig:experimental_results}. The qualitative correspondence between the measurements between the dislocated Gaussian and Laguerre decompositions makes it clear that we are, in fact, producing equivalent beams in the fundamental frequency.

Moving to the second harmonic results, in the case of the Laguerre decomposition with $\tau = \pm 1$, we can clearly observe the predicted dipin intensity at the origin, which is absent in the corresponding dislocated Gaussian decomposition. The absence of the dip for $\tau = 0$, even in the case of the Laguerre decomposition, also agrees with equation \eqref{eq:gamma_shg_peak}. Finally, in the second harmonic, we observe an overall decrease in the visibility of the fringes, which is consistent with a decrease in the coherence length, as predicted in \eqref{eq:shg_correlation_rp_continuous.}.
 
\section{Conclusion}
\label{sec:conclusion}
In this work, we analyzed the evolution of TGSM beams in a type-II SHG process. We found that different preparation methods can give rise to different output beams, depending on the SHG setup. We verified these predictions experimentally in the case of a single input beam with diagonal polarization.   In particular, an intensity dip at the origin is observed when the fundamental beam is prepared using a Laguerre-Gauss decomposition. We note that similar effects have been predicted for SHG where the input quantum fields display spatial antibunching \cite{nogueira01,caetano03,nogueira04}, and the SHG intensity can reach zero due to the photon statistics of the input field \cite{ether06}. However, in that case, the intensity minimum occurs for all $\mathbf{r}=\mathbf{r}^\prime$, not just at the origin.

The findings here presented reveal novel characteristics of TGSM beams that undergo a SHG process and sheds light on important, but often overlooked, details regarding this class of beams, such as the methods used to prepare them. Our results should be important in the study of the propagation of partially coherent light through nonlinear media.

\section*{Acknowledgments}
This work was funded by the Chilean agencies Fondo Nacional de Desarrollo Cient\'{i}fico y Tecnol\'{o}gico (FONDECYT - DOI 501100002850) (1230796, 1200266); National Agency of Research and Development (ANID) Millennium Science Initiative Program—ICN17-012; the Brazilian agencies
 Coordena\c c\~{a}o de Aperfei\c coamento de Pessoal de N\'\i vel Superior (CAPES DOI 501100002322), 
Funda\c c\~{a}o de Amparo \`{a} Pesquisa do Estado de Santa Catarina (FAPESC - DOI 501100005667),
Conselho Nacional de Desenvolvimento Cient\'{\i}fico e Tecnol\'ogico (CNPq - DOI 501100003593), 
Instituto Nacional de Ci\^encia e Tecnologia de Informa\c c\~ao Qu\^antica (INCT-IQ 465469/2014-0), Fundação de Amparo à Pesquisa do Estado do Rio de Janeiro (FAPERJ-CNE E-26/201.108/2021), and Fundação de Amparo à Pesquisa do Estado de São Paulo (FAPESP, Processo No. 2021/06823-5).

\bibliographystyle{elsarticle-num} 
\bibliography{refs}

\end{document}